\documentclass[pdflatex,sn-mathphys]{sn-jnl}
\jyear{2023}

\usepackage{caption}
\usepackage{multibib}
\newcites{supp}{Methods References}

\usepackage{afterpage}
\usepackage{rotating}
\usepackage{amsmath}
\newcommand{\angstrom}{\text{\normalfont\AA}}

\begin{document}

\title[A massive quiescent galaxy at redshift 4.658]{A massive quiescent galaxy at redshift 4.658}

\author[1]{\fnm{Adam C.} \sur{Carnall}}

\author[1]{\fnm{Ross J.} \sur{McLure}}

\author[1]{\fnm{James S.} \sur{Dunlop}}

\author[1]{\fnm{Derek J.} \sur{McLeod}}

\author[2]{\fnm{Vivienne} \sur{Wild}}

\author[1]{\fnm{Fergus} \sur{Cullen}}

\author[3]{\fnm{Dan} \sur{Magee}}

\author[1]{\fnm{Ryan} \sur{Begley}}

\author[4,5]{\fnm{Andrea} \sur{Cimatti}}

\author[1]{\fnm{Callum T.} \sur{Donnan}}

\author[1]{\fnm{Massissilia L.} \sur{Hamadouche}}

\author[1]{\fnm{Sophie M.} \sur{Jewell}}

\author[1]{\fnm{Sam} \sur{Walker}}

\affil[1]{\orgdiv{Institute for Astronomy, School of Physics \& Astronomy}, \orgname{University of Edinburgh, Royal Observatory}, \orgaddress{\city{Edinburgh}, \postcode{EH9 3HJ}, \country{UK}}}

\affil[2]{\orgdiv{School of Physics \& Astronomy}, \orgname{University of St Andrews}, \orgaddress{\street{North Haugh}, \city{St Andrews}, \postcode{KY16 9SS}, \country{UK}}}

\affil[3]{\orgdiv{Department of Astronomy and Astrophysics}, \orgname{UCO/Lick Observatory, University of California}, \orgaddress{\city{Santa Cruz}, \postcode{CA 95064}, \country{USA}}}

\affil[4]{\orgdiv{Department of Physics and Astronomy (DIFA)}, \orgname{University of Bologna}, \orgaddress{\postcode{Via Gobetti 93/2, I-40129}, \city{Bologna}, \country{Italy}}}

\affil[5]{\orgdiv{INAF}, \orgname{Osservatorio di Astrofisica e Scienza dello Spazio}, \orgaddress{\postcode{Via Piero Gobetti 93/3, I-40129}, \city{Bologna}, \country{Italy}}}

\maketitle
\vspace{-1cm}

\textbf{The extremely rapid assembly of the earliest galaxies during the first billion years of cosmic history is a major challenge for our understanding of galaxy formation physics \cite{Dunlop1996, Cimatti2004, Kriek2016, Schreiber2018, Girelli2019}. The advent of JWST has exacerbated this issue by confirming the existence of galaxies in significant numbers as early as the first few hundred million years \cite{Labbe2022, Donnan2022, Carnall2023b}. Perhaps even more surprisingly, in some galaxies, this initial highly efficient star formation rapidly shuts down, or quenches, giving rise to massive quiescent galaxies as little as 1.5 billion years after the Big Bang \cite{Glazebrook2017, Valentino2020}, however, due to their faintness and red colour, it has proven extremely challenging to learn about these extreme quiescent galaxies, or to confirm whether any exist at earlier times. Here we report the spectroscopic confirmation of a massive quiescent galaxy, GS-9209, at redshift, $\mathbf{z=4.658}$, just 1.25 billion years after the Big Bang, using JWST NIRSpec. From these data we infer a stellar mass of $\mathbf{M_* = 3.8\pm0.2\times10^{10}\ M_\odot}$, which formed over a $\mathbf{\simeq200}$ Myr period before this galaxy quenched its star formation activity at $\mathbf{z=6.5^{+0.2}_{-0.5}}$, when the Universe was $\mathbf{\simeq800}$ million years old. This galaxy is both a likely descendent of the highest-redshift submillimetre galaxies and quasars, and a likely progenitor for the dense, ancient cores of the most massive local galaxies.}

\begin{figure}
	\includegraphics[width=\columnwidth]{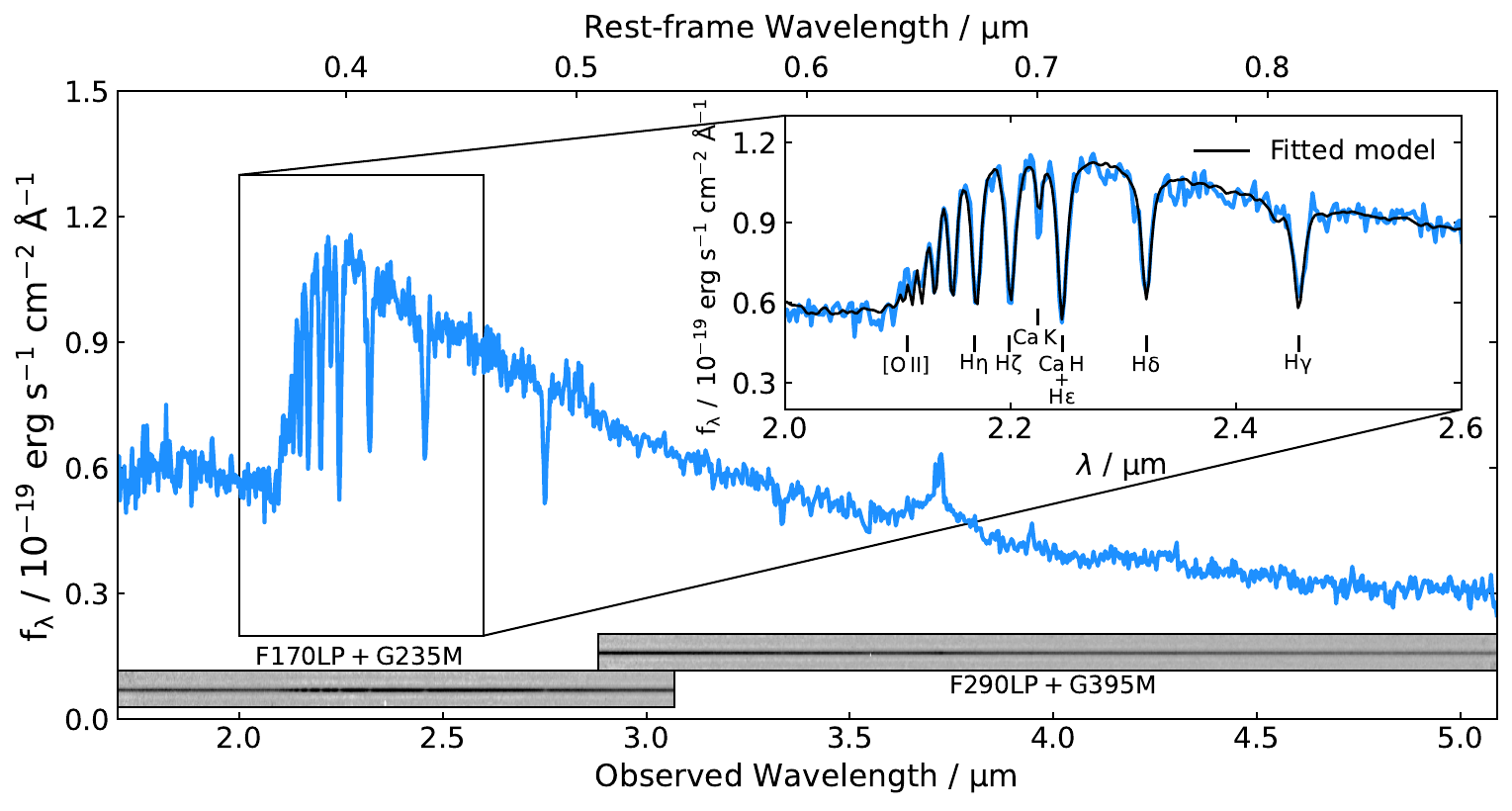}
    \caption{\textbf{JWST NIRSpec observations of GS-9209.} Data were taken on 16\textsuperscript{th} November 2022, using the G235M and G395M gratings ($R=1000$) with integration times of 3 hours and 2 hours respectively, providing wavelength coverage from $\lambda=1.7-5.1\mu$m. The galaxy is at a redshift of $z=4.6582\pm0.0002$, and exhibits extremely deep Balmer absorption lines. The spectrum strongly resembles that of an A-type star, and is reminiscent of lower-redshift post-starburst galaxies \cite{Goto2007, Wild2009, Wild2020}, clearly indicating this galaxy experienced a significant, rapid drop in star-formation rate (SFR) within the past few hundred million years. The spectral region from $\lambda=2.6-4.0\mu$m,  containing $H\beta$ and $H\alpha$, is shown at a larger scale in Figure \ref{fig:spectrum2}.}
    \label{fig:spectrum1}
\end{figure}

\begin{figure}
	\includegraphics[width=\columnwidth]{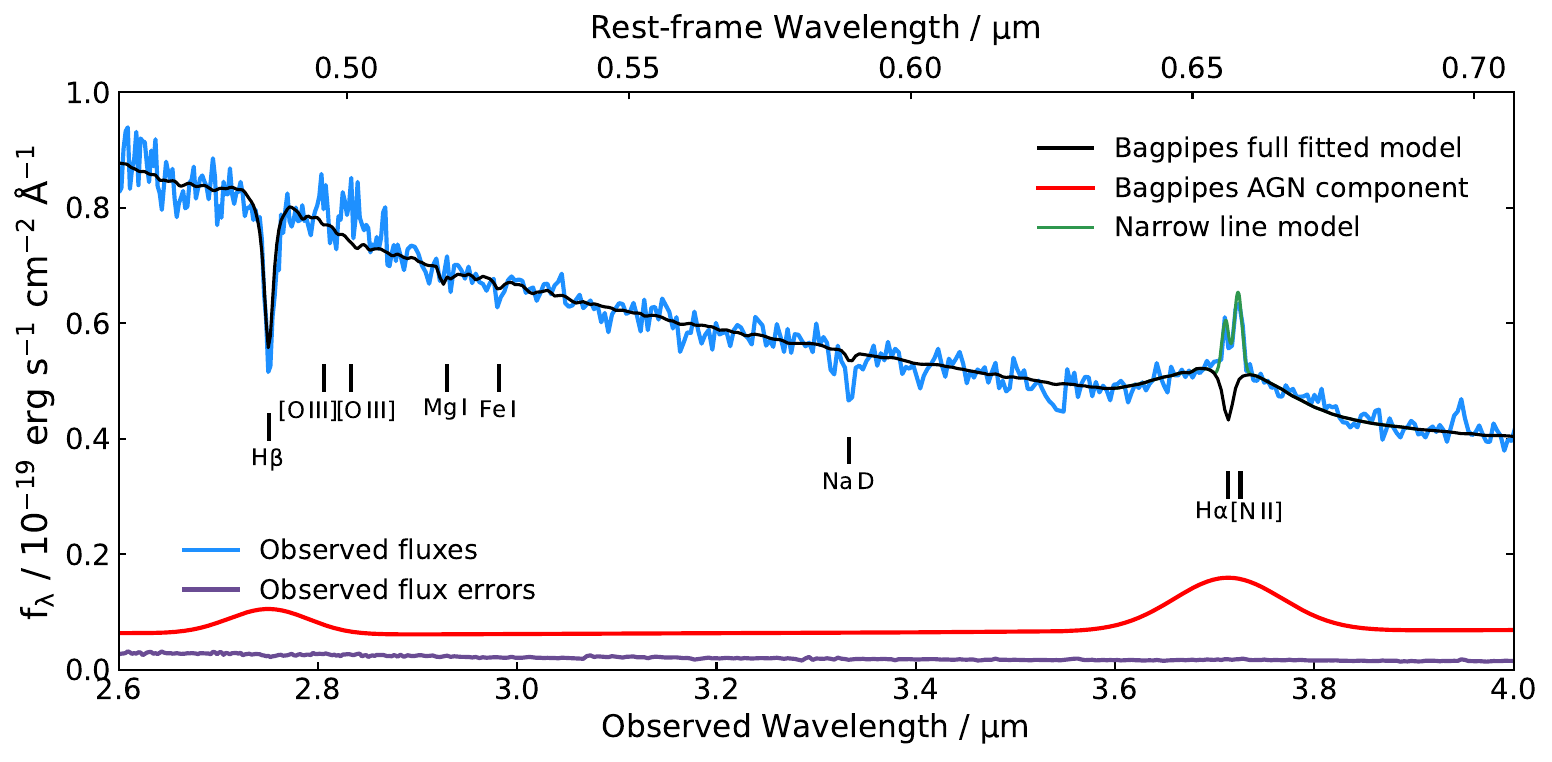}
    \caption{\textbf{JWST NIRSpec observations of GS-9209: zoom in on H$\boldsymbol{\beta}$ and H$\boldsymbol{\alpha}$}. Data are shown in blue, with their associated ($1\sigma$ standard deviation) uncertainties visible at the bottom in purple. The full Bagpipes fitted model is shown in black, with the AGN component shown in red. The narrow H$\alpha$ and [N\,\textsc{ii}] lines were masked during the Bagpipes fitting process, and subsequently fitted with Gaussian functions, shown in green. Key emission and absorption features are also marked.}
    \label{fig:spectrum2}
\end{figure}

During the past 5 years, several studies have identified GS-9209 as a candidate high-redshift massive quiescent galaxy \cite{Merlin2018, Carnall2020}, based on its blue colours at wavelengths, $\lambda = 2 - 8\ \mu$m and non-detection at millimetre wavelengths \cite{Santini2019}. GS-9209 is also not detected in X-rays \cite{Luo2017}, at radio wavelengths \cite{Bonzini2013}, or at $\lambda=24\ \mu$m \cite{Dunlop2007}. The faint, red nature of the source (with magnitudes $H_\mathrm{AB}=24.7$ and $K_\mathrm{AB}=23.6$) means that near-infrared spectroscopy with ground-based instrumentation is prohibitively expensive. The JWST NIRSpec data, shown in Figure \ref{fig:spectrum1}, reveal a full suite of extremely deep Balmer absorption features, with a H$\delta$ equivalent width (EW), as measured by the H$\delta_\mathrm{A}$ Lick index, of $7.9 \pm 0.3$ \AA, comparable to the most extreme values observed in the local Universe \cite{Kauffmann2003}. These spectral features strongly indicate this galaxy has undergone a sharp decline in star-formation rate (SFR) during the preceding few hundred Myr.

The spectrum exhibits only the merest suspicion of [O\,\textsc{ii}] 3727 \AA\ and [O\,\textsc{iii}] 4959\AA, 5007 \AA\ emission, and no apparent infilling of H$\beta$ or any of the higher-order Balmer absorption lines. However, as can be seen in Figure \ref{fig:spectrum2}, both H$\alpha$ and [N\textsc{ii}] 6584 \AA\ are clearly albeit weakly detected in emission, with H$\alpha$ also exhibiting an obvious broad component. This broad component, along with the relative strength of [N\,\textsc{ii}] compared with the narrow H$\alpha$ line, indicate the presence of an accreting supermassive black hole: an active galactic nucleus (AGN). However, the extreme EWs of the observed Balmer absorption features indicate that the continuum emission must be strongly dominated by the stellar component.

To measure the stellar population properties of GS-9209, we perform full spectrophotometric fitting using the Bagpipes code \cite{Carnall2018} (see Methods). Briefly, we first mask the wavelengths corresponding to [O\,\textsc{ii}], [O\,\textsc{iii}], narrow H$\alpha$ and [N\,\textsc{ii}], due to likely AGN contributions. We then fit a 22-parameter model for the stellar, dust, nebular and AGN components, as well as spectrophotometric calibration, to the spectroscopic data in combination with multi-wavelength photometry. Throughout the paper we report only statistical uncertainties on fitted parameters. It should be noted however that systematic uncertainties in galaxy spectral energy distribution analyses can be significantly larger \cite{Carnall2019a,Leja2019a,Pacifici2023}. We investigate the effect of our choice of star-formation history (SFH) model in the Methods section.

The resulting posterior median model is shown in black in Figs \ref{fig:spectrum1} and \ref{fig:spectrum2}. We obtain a stellar mass of log$_{10}(M_*/\rm{M}_\odot) = 10.58\pm0.02$, under the assumption of a Kroupa initial mass function (IMF) \cite{Kroupa2001}. We additionally recover a very low level of dust attenuation, with $A_V = 0.02\pm0.02$. The SFR we measure averaged over the past 100 Myr is consistent with zero, with a very stringent upper bound, though this is largely a result of our chosen SFH parameterisation \cite{Carnall2019a}. We provide a detailed discussion of the SFR of GS-9209 in Methods.

\begin{figure}
	\includegraphics[width=\columnwidth]{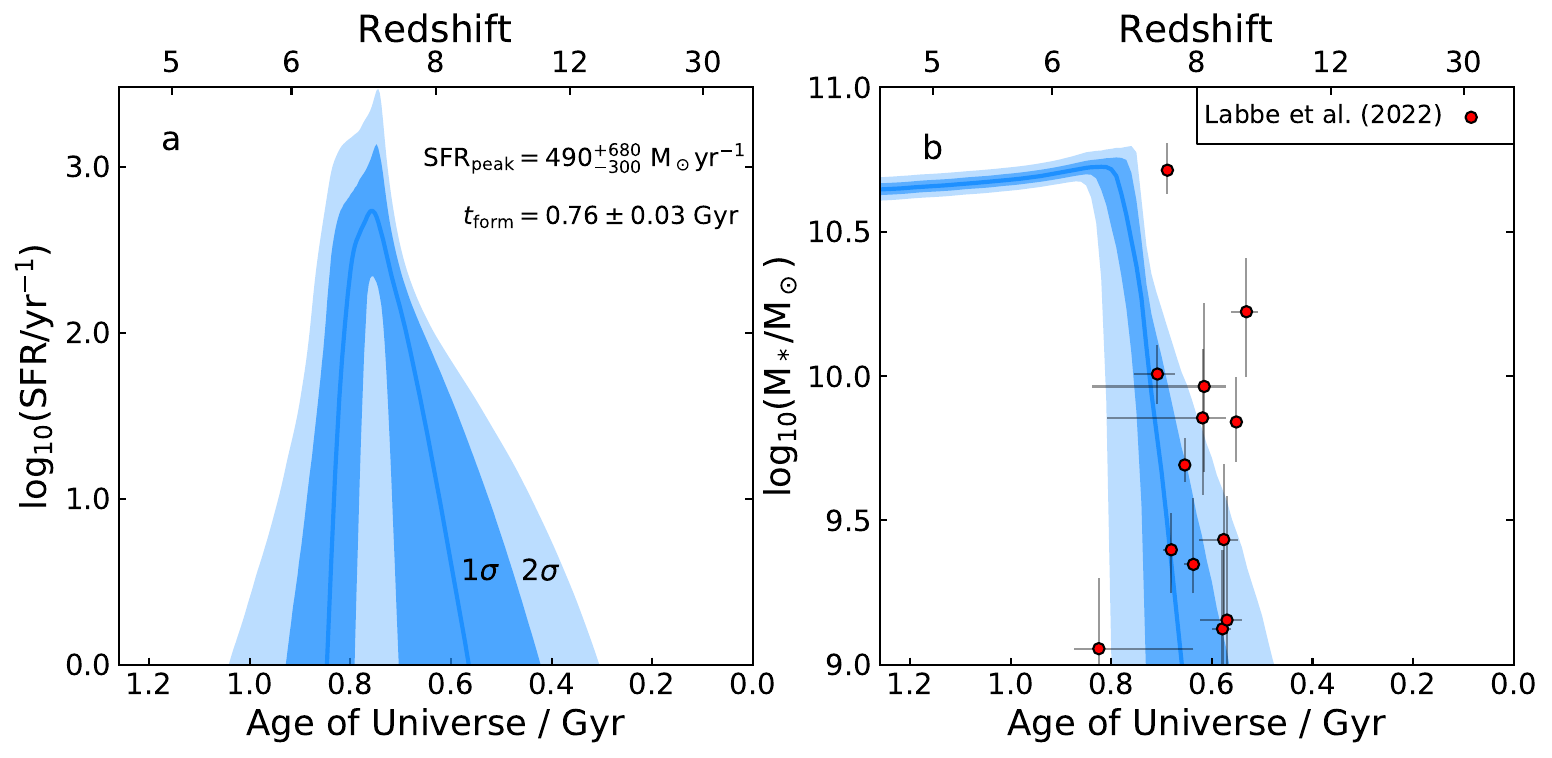}
    \caption{\textbf{The star-formation rate and stellar mass of GS-9209 as a function of time.} Panel a shows the star-formation rate (SFR) as a function of time (the star-formation history). Panel b shows the stellar mass as a function of time. The blue lines show the posterior medians, with the darker and lighter shaded regions showing the 1$\sigma$ and 2$\sigma$ confidence intervals respectively. We find a formation redshift, $z_\mathrm{form}=6.9\pm0.2$ and a quenching redshift, $z_\mathrm{quench} = 6.5^{+0.2}_{-0.5}$. The sample of massive $z\simeq8$ galaxy candidates from JWST CEERS reported by \cite{Labbe2022} is also shown in the right panel, demonstrating that these candidates are plausible progenitors for GS-9209. The uncertainties shown on the red points are $1\sigma$ standard deviation values.}
    \label{fig:sfh}
\end{figure}

The SFH we recover is shown in Figure \ref{fig:sfh}. We find that GS-9209 formed its stellar population largely during a $\simeq200$ Myr period, from around $600-800$ Myr after the Big Bang ($z\simeq7-8$). We recover a mass-weighted mean formation time, $t_\mathrm{form}=0.76\pm0.03$ Gyr after the Big Bang, corresponding to a formation redshift, $z_\mathrm{form}=6.9\pm0.2$. This is the redshift at which GS-9209 would have had half its current stellar mass, approximately log$_{10}(M_*/\mathrm{M_\odot}) = 10.3$. We find that GS-9209 quenched (which we define as the time at which its sSFR fell below 0.2 divided by the Hubble time, e.g., \cite{Pacifici2016}) at time $t_\mathrm{quench}=0.83^{+0.08}_{-0.06}$ Gyr after the Big Bang, corresponding to a quenching redshift, $z_\mathrm{quench} = 6.5^{+0.2}_{-0.5}$.

Our model predicts that the peak historical SFR for GS-9209 (at approximately $z_\mathrm{form}$) was within the range SFR$_\mathrm{peak} = 490^{+680}_{-300}$ M$_\odot$ yr$^{-1}$. This is similar to the SFRs of bright submillimetre galaxies (SMGs). The number density of SMGs with SFR $> 300$ M$_\odot$ yr$^{-1}$ at $5 < z < 6$ has been estimated to be $\simeq3\times10^{-6}$ Mpc$^{-3}$ \cite{Michalowski2017}. Extrapolation then suggests that the SMG number density at $z \simeq 7$ is $\simeq 1 \times 10^{-6}$ Mpc$^{-3}$, which  equates to $\simeq 1$ SMG at $z \simeq 7$ over the $\simeq 400$ square arcmin area from which GS-9209 and one other $z > 4$ quiescent galaxy were selected \cite{Carnall2020}. This broadly consistent number density suggests it is entirely plausible that GS-9209 went through a SMG phase at $z\simeq7$, shortly before quenching.

In panel b of Figure \ref{fig:sfh}, we show the positions of the massive, high-redshift galaxy candidates recently reported by \cite{Labbe2022} in the first imaging release from the JWST CEERS survey. The positions of these galaxies are broadly consistent with the SFH of GS-9209 at $z\simeq7-8$. It should however be noted that, as previously discussed, GS-9209 was selected as one of only two robustly identified $z > 4$ massive quiescent galaxies in an area roughly 10 times the size of the initial CEERS imaging area \cite{Carnall2020}. It therefore seems unlikely that a large fraction of the candidates reported by \cite{Labbe2022} will evolve in a similar way to GS-9209 over the redshift interval from $z\simeq5-8$.

From our Bagpipes full spectral fit, we measure an observed broad H$\alpha$ flux of $f_{\mathrm{H}\alpha\mathrm{,\, broad}} = 1.26 \pm 0.08 \times 10^{-17}= $ erg s$^{-1}$ cm$^{-2}$ and full width at half maximum (FWHM) of $10300\pm700$ km s$^{-1}$ in the rest frame. This line width, whilst very broad, is consistent with rest-frame UV broad line widths measured for some $z\simeq6$ quasars (e.g., \cite{Chehade2018, Onoue2019}).

As visualised in Figure \ref{fig:spectrum2}, we fit Gaussian components to the narrow H$\alpha$ and [N\,\textsc{ii}] lines following subtraction of our best-fitting Bagpipes model (see Methods). We obtain a H$\alpha$ narrow-line flux of $1.58\pm0.10\ \times\ 10^{-18}$ erg s$^{-1}$ cm$^{-2}$ and a [N\,\textsc{ii}] flux of $1.56\pm0.10\ \times\ 10^{-18}$ erg s$^{-1}$ cm$^{-2}$, giving a line ratio of log$_{10}$([N\,\textsc{ii}]/H$\alpha) = -0.01\pm0.04$. This line ratio is significantly higher than would be expected as a result of ongoing star formation, and is consistent with excitation due to an AGN or shocks resulting from galactic outflows \cite{Kewley2013}. Such outflows are commonly observed in post-starburst galaxies at $z\gtrsim1$ \cite{Maltby2019}. We discuss what can be learned about the star-formation rate of GS-9209 from the observed H$\alpha$ flux in Methods.

We estimate the black-hole mass for GS-9209, $M_\bullet$, from our combined H$\alpha$ flux and broad-line width, using the relation presented in Equation 6 of \cite{Greene2005}, obtaining log$_{10}(M_\bullet/$M$_\odot)= 8.7\pm0.1$. From our Bagpipes full spectral fit, we infer a stellar velocity dispersion, $\sigma=247\pm16$ km s$^{-1}$ for GS-9209, after correcting for the intrinsic dispersion of our template set, as well as instrumental dispersion. Given this measurement, the relationship between velocity dispersion and black-hole mass presented by \cite{Kormendy2013} predicts log$_{10}(M_\bullet/$M$_\odot)= 8.9\pm0.1$.

Given the broad agreement between these estimators, it seems reasonable to conclude that GS-9209 contains a supermassive black hole with a mass of approximately half a billion to a billion Solar masses. It is interesting to note that this is $\simeq4-5$ times the black-hole mass that would be expected given the stellar mass of the galaxy, assuming this is equivalent to the bulge mass. This is consistent with the observed increase in the average black-hole to bulge mass ratio for massive galaxies from $0 < z < 2$ \cite{McLure2006}. The large amount of historical AGN accretion implied by this significant black hole mass suggests that AGN feedback may have been responsible for quenching this galaxy \cite{Maiolino2012}.

\begin{figure}
	\includegraphics[width=\columnwidth]{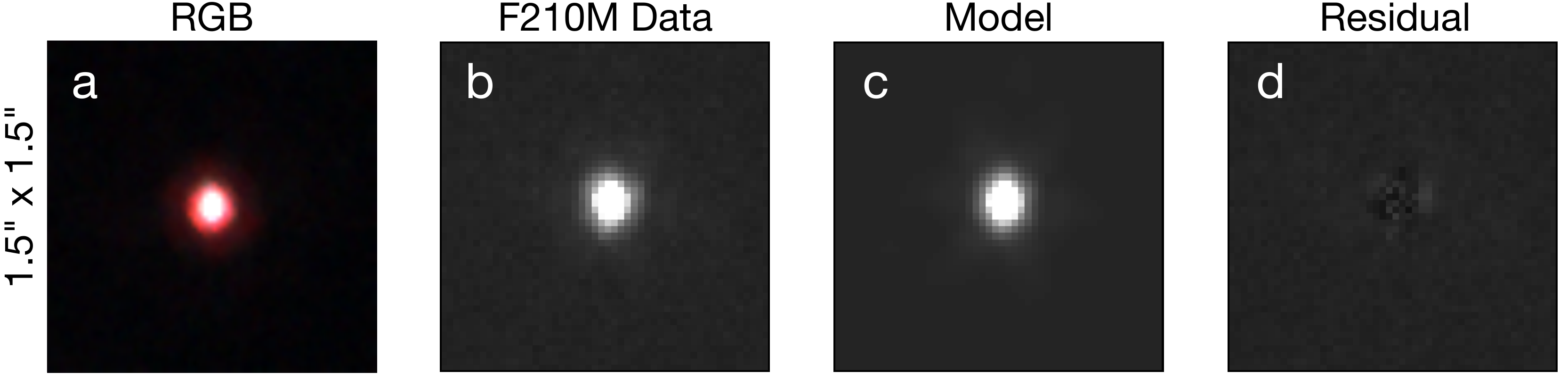}
    \caption{\textbf{JWST NIRCam imaging of GS-9209.} Each cutout image shows an area of $1.5^{\prime\prime}\times1.5^{\prime\prime}$. Panel a is a RGB image, constructed with F430M as red, F210M as green and F182M as blue. Panel b shows the F210M image, with our posterior median PetroFit model shown in Panel c. Panel d shows the residuals between model and data, on the same colour scale as panels b and c.}
    \label{fig:images}
\end{figure}

GS-9209 is an extremely compact source, which is only marginally resolved in the highest-resolution available imaging data. We measure the size of GS-9209 using newly available JWST NIRCam F210M-band imaging, which has a FWHM of $\simeq0.07^{\prime\prime}$ (see Methods). Accounting for the AGN point-source contribution, we measure an effective radius, $r_e=0.033\pm0.003^{\prime\prime}$ for the stellar component of GS-9209, along with a S\'ersic index, $n=2.3\pm0.3$. At $z=4.658$, this corresponds to $r_e=215\pm20$ parsecs.  This is consistent with previous measurements by the CANDELS/3DHST team \cite{vanderwel2014}, and is $\simeq0.7$ dex below the mean relationship between $r_e$ and stellar mass for quiescent galaxies at $z\simeq1$ \cite{vanderwel2014, Hamadouche2022}. This is interesting given that post-starburst galaxies $z\simeq1$ are known to be more compact than is typical for the wider quiescent population \cite{Almaini2017}. We calculate a stellar-mass surface density within $r_e$ of log$_{10}(\Sigma_\mathrm{eff}/$M$_\odot$ kpc$^{-2}) = 11.15\pm0.08$, consistent with the densest stellar systems in the Universe \cite{Hopkins2010}. We show the F210M data for GS-9209, along with our posterior-median model in Figure \ref{fig:images}.

We estimate the dynamical mass using our size and velocity dispersion measurements (e.g., \cite{Maltby2019}), obtaining a value of log$_{10}(M_\mathrm{dyn}/$M$_\odot)= 10.3\pm0.1$. This is $\simeq0.3$ dex lower than the stellar mass we measure. As GS-9209 is only marginally resolved, even in JWST imaging data, and due to the presence of the AGN component, it is plausible that our measured $r_e$ may be subject to systematic uncertainties. Furthermore, since the pixel scale of NIRSpec is 0.1$^{\prime\prime}$, our velocity dispersion measurement may not accurately represent the central velocity dispersion, leading to an underestimated dynamical mass. It should also be noted that the stellar mass we measure is strongly dependent on our assumed IMF. A final, intriguing possibility would be a high level of rotational support in GS-9209, as has been observed for quiescent galaxies at $2 < z < 3$ \cite{newman2018}. Unfortunately, the extremely compact nature of the source makes any attempt at resolved studies extremely challenging, even with the JWST NIRSpec integral field unit. Resolved kinematics for this galaxy would be a clear use case for the High Angular Resolution Monolithic Optical and Near-infrared Integral field spectrograph (HARMONI) planned for the Extremely Large Telescope (ELT).

GS-9209 demonstrates unambiguously that massive galaxy formation was already well underway within the first billion years of cosmic history and that the earliest onset of galaxy quenching was no later than $\simeq800$ Myr after the Big Bang. Based on the properties we measure, GS-9209 seems likely to be associated with the most extreme galaxy populations currently known at $z>5$, such as the highest-redshift submillimetre galaxies and quasars (e.g., \cite{Decarli2017, Onoue2019, Riechers2021}). GS-9209 and similar objects (e.g., \cite{Carnall2023b}) are also likely progenitors for the dense, ancient cores of the most massive galaxies in the local Universe.


\newpage

\captionsetup[table]{name=Extended Data Table}
\captionsetup[figure]{name=Extended Data Figure}

\section*{Methods}\label{sec:method}

\subsection*{Spectroscopic data and reduction}

The spectroscopic data shown in Figure \ref{fig:spectrum1} were taken on 16\textsuperscript{th} November 2022. The target was acquired directly via Wide Aperture Target Acquisition (WATA), meaning the object is extremely well centred. Spectroscopic observations were taken through the NIRSpec fixed slit (S200A1), which has a width of 0.2$^{\prime\prime}$. Data were taken using the G235M and G395M gratings, providing an average spectral resolution of $R=1000$. With each grating, a total of 5 integrations were taken at different dither positions along the slit. The readout pattern used was NRSIRS2, with 30 and 20 groups per integration for the two gratings respectively, providing total integration times of 3 hours and 2 hours respectively.

We reduce our NIRSpec data with the JWST Science Calibration Pipeline v1.8.4, using version 1017 of the JWST calibration reference data. To improve the spectrophotometric calibration of our data, we also reduce observations of the A-type standard star 2MASS J18083474+6927286 \citesupp{Gordon2022}, taken as part of JWST commissioning programme 1128 (PI: L{\"u}tzgendorf) \citesupp{Luetzgendorf2022} using the same instrument modes. We compare the resulting stellar spectrum against a spectral model for this star from the CALSPEC library \citesupp{Bohlin2014} to construct a calibration function, which we then apply to our observations of GS-9209. The resulting spectrophotometry is well matched with the available near-infrared photometric data, and the calibration polynomial we fit along with our Bagpipes model only results in further calibration changes at the $\simeq10$ per cent level. We additionally find that the uncertainties output by the pipeline are only moderately underestimated, with the errorbar expansion term in our Bagpipes model resulting in an increase of 50 per cent to the pipeline-produced uncertainties, in agreement with other recent analyses\footnote{e.g., \url{https://github.com/spacetelescope/jwst/issues/7362}}.

\subsection*{Photometric data reduction}

The majority of our photometric data are taken directly from the CANDELS GOODS South catalogue \citesupp{Guo2013}. We supplement this with new JWST NIRCam photometric data taken as part of the Ultra Deep Field Medium-Band Survey \citesupp{Williams2023} (Programme ID: 1963; PI: Williams). Data are available in the F182M, F210M, F430M, F460M and F480M bands. We reduce these data using the PRIMER Enhanced NIRCam Image-processing Library (PENCIL, e.g., \cite{Donnan2022}), a custom version of the JWST Science Calibration Pipeline (v1.8.0), and using version 1011 of the JWST calibration reference data. We measure photometric fluxes for GS-9209 in large, 1$^{\prime\prime}$-diameter apertures to ensure we measure the total flux in each band (the object is isolated, with no other sources within this radius, see Figure \ref{fig:images}). We measure uncertainties as the standard deviation of flux values in the nearest 100 blank-sky apertures, masking out nearby objects (e.g., \citesupp{McLeod2016}).

\subsection*{Bagpipes full spectral fitting}\label{sec:method_fitting}

We fit the available photometry in parallel with our new spectroscopic data using the Bagpipes code \cite{Carnall2018}. Our model has a total of 22 free parameters, describing the stellar, dust, nebular and AGN components of the spectrum. A full list of these parameters, along with their associated priors, is given in Extended Data Table \ref{table:params}. We fit our model to the data using the MultiNest nested sampling algorithm \citesupp{Skilling2006, Buchner2014, Feroz2019}. The full Bagpipes fit to our combined dataset, along with residuals, is shown in Extended Data Figure \ref{fig:spectrum3}. Posterior percentiles for our fit to the data are given in Extended Data Table \ref{table:params_values}.

We use the 2016 updated version of the BC03 \citesupp{Bruzual2003, Chevallard2016} stellar population models, using the MILES stellar spectral library \citesupp{Sanchez-Blazquez2006} and updated stellar evolutionary tracks \citesupp{Bressan2012, Marigo2013}. We assume a double-power-law star-formation-history model (e.g., \cite{Carnall2018, Carnall2019a}). We allow the logarithm of the stellar metallicity, $Z_*$ to vary freely from log$_{10}(Z_*/\mathrm{Z_\odot}) = -2.45$ to 0.55. These are the limits of the range spanned by the BC03 model grid relative to our adopted Solar metallicity value ($\mathrm{Z_\odot} = 0.0142$\ \citesupp{Asplund2009}).

We mask out the narrow emission lines in our spectrum during our Bagpipes fitting due to likely AGN contributions, whereas Bagpipes is only capable of modelling emission lines from star-forming regions. We do however still include a nebular model in our Bagpipes fit to allow for the possibility of nebular continuum emission from star-forming regions. We assume a stellar-birth-cloud lifetime of 10 Myr, and vary the logarithm of the ionization parameter, U, from log$_{10}(U) = -4$ to $-2$. We also allow the logarithm of the gas-phase metallicity, $Z_\mathrm{g}$, to vary freely from log$_{10}(Z_\mathrm{g}/\mathrm{Z_\odot}) = -2.45$ to 0.55. Because our eventual fitted model only includes an extremely small amount of star formation within the last 10 Myr for GS-9209, this nebular component makes a negligible contribution to the fitted model spectrum.

We model attenuation of the above components by dust using the model of \citesupp{Noll2009, Salim2018}, which is parameterised as a power-law deviation from the Calzetti dust attenuation law \citesupp{Calzetti2000}, and also includes a Drude profile to model the 2175\AA\ bump. We allow the $V-$band attenuation, $A_V$ to vary from $0-4$ magnitudes. We further assume that attenuation is multiplied by an additional factor for all stars with ages below 10 Myr, and resulting nebular emission. This factor is commonly assumed to be 2, however we allow this to vary from 1 to 5.

We allow redshift to vary, using a narrow Gaussian prior with a mean of 4.66 and standard deviation of 0.01. We additionally convolve the spectral model with a Gaussian kernel in velocity space, to account for velocity dispersion in our target galaxy. The width of this kernel is allowed to vary with a logarithmic prior across a range from $50-500$ km s$^{-1}$. The resolution of our spectroscopic data is high enough that the total dispersion is dominated by stellar velocity dispersion within the target galaxy, which has a standard deviation of $\sigma\simeq250$ km s$^{-1}$, compared with the average instrumental dispersion of $\sigma\simeq128$  km s$^{-1}$.

Separately from the above components, we also include a model for AGN continuum, broad H$\alpha$ and H$\beta$ emission. Following \citesupp{vandenberk2001}, we model AGN continuum emission with a broken power law, with two spectral indices and a break at $\lambda_\mathrm{rest}=5000$ \AA\ in the rest frame. We vary the spectral index at $\lambda_\mathrm{rest} < 5000$ \AA\ using a Gaussian prior with a mean value of $\alpha_\lambda=-1.5$ ($\alpha_\nu=-0.5)$ and standard deviation of 0.5. We also vary the spectral index at $\lambda_\mathrm{rest} > 5000$ \AA\ using a Gaussian prior with a mean value of $\alpha_\lambda=0.5$ ($\alpha_\nu=-2.5)$ and standard deviation of 0.5. We parameterise the normalisation of the AGN continuum component using $f_{5100}$, the flux at rest-frame 5100 \AA, which we allow to vary with a linear prior from 0 to $10^{-19}$ erg s$^{-1}$ cm$^{-2}$ \AA$^{-1}$.

We model broad H$\alpha$ with a Gaussian component, varying the normalisation from 0 to $2.5 \times 10^{-17}$ erg s$^{-1}$ cm$^{-2}$ using a linear prior, and the velocity dispersion from $1000-5000$ km s$^{-1}$ in the rest frame using a logarithmic prior. We also include a broad H$\beta$ component in the model, which has the same parameters as the broad H$\alpha$ line, but with normalisation divided by the standard 2.86 ratio from Case B recombination theory. However, as shown in Figure \ref{fig:spectrum2}, this H$\beta$ model peaks at around the noise level in our spectrum, and the line is therefore plausible in not being obviously detected in the observed spectrum.

We include intergalactic medium (IGM) absorption using the model of \citesupp{Inoue2014}. To allow for imperfect spectrophotometric calibration of our spectroscopic data, we also include a second-order Chebyshev polynomial (e.g., \citesupp{Carnall2019b, Johnson2021, Carnall2022}), which the above components of our combined model are all divided by before being compared with our spectroscopic data. We finally fit an additional white noise term, which multiplies the spectroscopic uncertainties from the JWST pipeline by a factor, $a$, which we vary with a logarithmic prior from $1-10$.

\subsection*{Investigation of alternative star-formation history models}

The functional forms used to model galaxy star-formation histories are well known to significantly affect physical parameter inferences \cite{Carnall2019a, Leja2019a, Pacifici2023}, with the degree of systematic uncertainty highly dependent on the physical parameter of interest, the type of data, and the galaxy being studied. In this section, we test the dependence of our inferred formation and quenching times for GS-9209 on the SFH model used. We re-run our Bagpipes full-spectral-fitting analysis, substituting the double-power-law SFH model described above, firstly for the continuity non-parametric model \cite{Leja2019a}, and secondly for a simple top-hat (constant) SFH model. For the continuity model, we use 8 time bins, with bin edges at 0, 10, 100, 200, 400, 600, 800, 1000 and 1260 Myr before observation. For the top-hat model, we vary the time at which star-formation turned on with a uniform prior between the Big Bang and time of observation. We vary the time at which star formation then stopped with a uniform prior from the time at which star formation turned on to the time of observation.

The results of these alternative fitting runs are shown in Extended Data Figure \ref{fig:alt_sfhs}. This figure shows two alternative versions of Figure \ref{fig:sfh}, with the continuity non-parametric model results shown in panels a and b, and the top-hat model results shown in panels c and d. The SFH posteriors shown, whilst varying in their detailed shapes, are in good overall agreement with our original double-power-law model. For the double-power-law model, we recover $t_\mathrm{form}=0.76\pm0.03$ Gyr and $t_\mathrm{quench}=0.83^{+0.08}_{-0.06}$ Gyr after the Big Bang. The values returned under the assumption of these other two models are consistent to within 1$\sigma$. For the continuity non-parametric model we recover $t_\mathrm{form}=0.74^{+0.02}_{-0.03}$ Gyr and $t_\mathrm{quench}=0.86^{+0.19}_{-0.01}$ Gyr. For the top-hat model we recover $t_\mathrm{form}=0.74\pm0.02$ Gyr and $t_\mathrm{quench}=0.91^{+0.04}_{-0.06}$ Gyr. Both of these models also produce stronger constraints on the peak historical SFR of GS-9209 at a lower level than the double-power-law model, though still consistent within 1$\sigma$. We conclude that our key results are not strongly dependent on our choice of star-formation history model.

\subsection*{AGN contribution and fitting of narrow emission lines}

From our Bagpipes full spectral fit, we recover an observed AGN continuum flux at rest-frame wavelength, $\lambda_\mathrm{rest} = 5100$ \AA\ of $f_{5100} = 0.06\pm0.01 \times 10^{-19}$ erg s$^{-1}$ cm$^{-2}$ \AA$^{-1}$. This is approximately 7.5 per cent of the total observed flux from GS-9209 at $\lambda=2.9\mu$m. We measure a power-law index for the AGN continuum emission of $\alpha_\lambda = -0.5\pm0.3$ at $\lambda_\mathrm{rest} < 5000$ \AA, and $\alpha_\lambda = 0.4\pm0.3$ at $\lambda_\mathrm{rest} > 5000$ \AA. The AGN contribution to the continuum flux from GS-9209 rises to $\simeq10$ per cent at the blue end of our spectrum ($\lambda=1.7\mu$m), and $\simeq20$ per cent at the red end ($\lambda=5\mu$m). Just above the Lyman break at $\lambda\simeq7000\,$\AA, the AGN contribution is $\simeq35$ per cent of the observed flux.

Given our measured $f_{\mathrm{H}\alpha\mathrm{,\,broad}}$, which is more direct than our AGN continuum measurement, the average relation for local AGN presented by \cite{Greene2005} predicts $f_{5100}$ to be $\simeq0.2$ dex brighter than we measure. However, given the intrinsic scatter of 0.2 dex they report, our measured $f_{5100}$ is only 1$\sigma$ below the mean relation. The extreme equivalent widths of the observed Balmer absorption features firmly disfavour stronger AGN continuum emission.

We fit the narrow H$\alpha$ and [N\,\textsc{ii}] lines in our spectrum as follows. We first subtract from our observed spectrum the posterior median Bagpipes model from our full spectral fitting. We then simultaneously fit Gaussian components to both lines, assuming the same velocity width for both, which is allowed to vary. This process is visualised in Figure \ref{fig:spectrum2}. We also show the broad H$\beta$ line in our AGN model, for which we assume the same width as broad H$\alpha$, as well as Case B recombination. It can be seen that the broad H$\beta$ line peaks at around the noise level in our spectrum, and is hence too weak to be clearly observed in our data.

\subsection*{The star-formation rate of GS-9209}

In this section we discuss the available observational indicators for the SFR of GS-9209. The commonly applied sSFR threshold for defining quiescent galaxies is sSFR$_\mathrm{threshold} = 0.2/t_\mathrm{H}$, where $t_\mathrm{H}$ is the age of the Universe \cite{Pacifici2016}. For GS-9209 at $z=4.658$ and log$_{10}(M_*/$M$_\odot) \simeq 10.6$, this corresponds to log$_{10}($sSFR$_\mathrm{threshold}/$yr$^{-1}) \simeq -9.8$, or SFR$_\mathrm{threshold} \simeq 6$ M$_\odot$ yr$^{-1}$.

In \citesupp{Santini2021}, the authors report that GS-9209 is undetected in ALMA band 6 data, with a flux of $-0.05\pm0.16$ mJy/beam, from which they derive a $1\sigma$ upper limit on SFR of 41 M$_\odot$ yr$^{-1}$. They also perform a stacking experiment, with stacked ALMA band 6 data for a sample of 20 objects selected as $3 < z < 5$ quiescent galaxies (including GS-9209) still yielding no detection, implying the average SFR for this sample is significantly below the individual-object detection limit. The extremely blue spectral shape of this object in the rest-frame red-optical to near-infrared (observed-frame $2-8\,\mu$m, see Extended Data Figure \ref{fig:spectrum3}) is also consistent with no significant obscured star-forming or AGN component. Deeper ALMA data for this object would be of value for setting a more stringent direct upper bound on obscured star formation.

As discussed in the main text, the high [N\,\textsc{ii}]/H$\alpha$ ratio in our observed spectrum strongly suggests that this line emission is powered by AGN activity or shocks. However, if we assume all the narrow H$\alpha$ emission is driven by ongoing star formation, neglecting dust attenuation we obtain SFR = $1.9\pm0.1$ M$_\odot$ yr$^{-1}$ \citesupp{Kennicutt2012}, corresponding to log$_{10}($sSFR/yr$^{-1})=-10.3\pm0.1$. Measurements of the average dust attenuation on H$\alpha$ emission, $A_{\mathrm{H}\alpha}$, are not yet available at $z\simeq5$, however, from $0 < z < 2$, stellar mass is found to be the most important factor in predicting the level of dust attenuation (e.g., \citesupp{Garn2010, Shapley2022}), with little evolution observed across this redshift interval. At $z\simeq2.3$, the average $A_{\mathrm{H}\alpha}$ for galaxies with log$_{10}(M_*/$M$_\odot) \simeq 10.6$ is 1.25 magnitudes \citesupp{Shapley2022}, which would suggest that the SFR of GS-9209 is roughly 6 M$_\odot$ yr$^{-1}$. However, given the multiple lines of evidence we uncover for a significant non-stellar component to the H$\alpha$ line, combined with the fact that the extremely low stellar continuum $A_V$ implies the gas-phase attenuation is also low \citesupp{Sanders2021}, it is likely that the sSFR of GS-9209 is considerably lower than the threshold normally applied for selecting quiescent galaxies.

\subsection*{Size measurement from F210M-band imaging}\label{sec:method_size}

The CANDELS/3DHST team \cite{vanderwel2014} measured an effective radius, $r_e = 0.029\pm0.002^{\prime\prime}$ for GS-9209 in the HST F125W filter via S\'ersic fitting, along with a S\'ersic index, $n=6.0\pm0.8$. At $z=4.658$, this corresponds to $r_e=189\pm13$ parsecs. We update this measurement using the newly acquired JWST NIRCam F210M imaging data discussed above. We model the light distribution of GS-9209 using PetroFit \citesupp{Geda2022}. We fit these PetroFit models to our data using the MultiNest nested sampling algorithm \citesupp{Skilling2006, Buchner2014, Feroz2019}. We use F210M in preference to the F182M band due to the smaller AGN contribution in F210M and the fact that it sits above the Balmer break, therefore being more representative of the stellar mass present rather than any ongoing star formation.

As our spectroscopic data contains strong evidence for an AGN, we fit both S\'ersic and delta-function components simultaneously, convolved by an empirically estimated PSF, derived by stacking bright stars. In preliminary fitting, we find that the relative fluxes of these two components are entirely degenerate with the S\'ersic parameters. We therefore predict the AGN contribution to the flux in this band based on our full-spectral-fitting result, obtaining a value of $8\pm1$ per cent. We then impose this as a Gaussian prior on the relative contributions from the S\'ersic and delta function components. The 11 free parameters of our model are the overall flux normalisation, which we fit with a logarithmic prior, the effective radius, $r_e$, S\'ersic index, $n$, ellipticity and position angle of the S\'ersic component, the x and y centroids of both components, the position angle of the point spread function, and the fraction of light in the delta-function component, which we fit with a Gaussian prior with a mean of 8 per cent and standard deviation of 1 per cent, based on our full spectral fitting result.

Deeper imaging data in the F200W or F277W bands (e.g., from the JWST Advanced Deep Extragalactic Survey; JADES) will provide a useful check on our size measurement for GS-9209, particularly given the lower AGN fraction in the F277W band.

\section*{Data Availability}

The datasets analysed during the current study are available from the Mikulski Archive for Space Telescopes (MAST) repository at \url{https://mast.stsci.edu}. The spectrum for GS-9209 was observed under JWST Programme ID 2285 (PI: Carnall). This programme has a 12-month proprietary period, and data will automatically become publicly available via MAST on 16\textsuperscript{th} November 2023. Reduced data products are available from the corresponding author upon request.

\section*{Code Availability}

The Bagpipes code is publicly available at \url{https://github.com/ACCarnall/bagpipes}. The PetroFit code is publicly available at \url{https://github.com/PetroFit/petrofit}. The JWST data reduction pipeline is publicly available at \url{https://github.com/spacetelescope/jwst}.


\section*{Acknowledgements}

The authors would like to thank James Aird for helpful discussions. A.\,C. Carnall thanks the Leverhulme Trust for their support via a Leverhulme Early Career Fellowship. R.\,J. McLure, J.\,S.~Dunlop, D.\,J. McLeod,  V. Wild, R. Begley, C.\,T. Donnan and M.\,L. Hamadouche acknowledge the support of the Science and Technology Facilities Council. F. Cullen acknowledges support from a UKRI Frontier Research Guarantee Grant (grant reference EP/X021025/1). A. Cimatti acknowledges support from the grant PRIN MIUR 2017 - 20173ML3WW\_001.

\section*{Author Contributions}

ACC led the preparation of the observing proposal, reduction and analysis of the data, and preparation of the manuscript. RJM, JSD, VW, FC and AC provided advice and assistance with data reduction, analysis and interpretation, as well as consulting on the preparation of the observing proposal. DJM, DM, RB and CTD reduced the JWST imaging data and prepared the empirical PSF. DJM, MLH and SMJ assisted with measurement of the size and morphology of GS-9209. SW assisted with selection of GS-9209 from the CANDELS catalogues prior to the observing proposal being submitted. All authors assisted with preparation of the final published manuscript.

\section*{Author Information}

The authors declare that they have no competing financial interests. Correspondence and requests for materials should be addressed to ACC (email: adam.carnall@ed.ac.uk).

\afterpage{
\begin{sidewaystable*}
\caption{The 22 free parameters of the Bagpipes model we fit to our spectroscopic and photometric data, along with their associated prior distributions. The upper limit on $\tau$, $t_\mathrm{obs}$, is the age of the Universe as a function of redshift. Logarithmic priors are all applied in base ten. For parameters with Gaussian priors, the mean is $\mu$ and the standard deviation is $\sigma$.}
\begingroup
\setlength{\tabcolsep}{3pt}
\renewcommand{\arraystretch}{1.1}
\begin{tabular}{llllllll}
\hline
Component & Parameter & Symbol / Unit & Range & Prior & \multicolumn{2}{l}{Hyper-parameters} \\
\hline
General & Redshift & $z$ & (4.6, 4.7) & Gaussian & $\mu=4.66$ & $\sigma=0.01$ \\
& Stellar velocity dispersion & $\sigma$ / km s$^{-1}$ & (50, 500) & Logarithmic & & \\
\hline
SFH & Total stellar mass formed & $M_*\ /\ \mathrm{M_\odot}$ & (1, $10^{13}$) & Logarithmic & & \\
& Stellar metallicity & $Z_*\ /\ \mathrm{Z_\odot}$ & (0.00355, 3.55) & Logarithmic & & \\
& Double-power-law falling slope & $\alpha$ & (0.01, 1000) & Logarithmic & & \\
& Double-power-law rising slope & $\beta$ & (0.01, 1000) & Logarithmic & & \\
& Double-power-law turnover time & $\tau$ / Gyr & (0.1, $t_\mathrm{obs}$) & Uniform & & \\
\hline
Dust & $V-$band attenuation & $A_V$ / mag & (0, 4) & Uniform & & \\
& Deviation from Calzetti slope & $\delta$ & ($-0.3$, 0.3) & Gaussian & $\mu = 0$ & $\sigma$ = 0.1 \\
& Strength of 2175\AA\ bump & $B$ & (0, 5) & Uniform & & \\
& Attenuation ratio for birth clouds & $\epsilon$ & (1, 5) & Uniform &  &  \\
\hline

AGN & Power law slope ($\lambda < 5000$ \AA) & $\alpha_{\lambda < 5000\mathrm{\angstrom}}$ & ($-2$, $2$) & Gaussian & $\mu=-1.5$ & $\sigma=0.5$ \\
& Power law slope ($\lambda > 5000$ \AA) & $\alpha_{\lambda > 5000\mathrm{\angstrom}}$ & ($-2$, $2$) & Gaussian & $\mu=0.5$ & $\sigma=0.5$ \\
& H$\alpha$ broad-line flux & $f_{\mathrm{H}\alpha\mathrm{,\,broad}}$ / erg s$^{-1}$ cm$^{-2}$ & (0, $2.5\times10^{-17}$) & Uniform & & \\
& H$\alpha$ broad-line velocity dispersion & $\sigma_{\mathrm{H}\alpha\mathrm{,\,broad}}$ / km s$^{-1}$ & (1000, 5000) & Logarithmic & & \\
& Continuum flux at $\lambda=5100$ \AA & $f_{5100}$ / erg s$^{-1}$ cm$^{-2}$ \AA$^{-1}$ & (0, $10^{-19}$) & Uniform & & \\

\hline

Nebular & Ionization parameter & $U$ & ($10^{-4}$, $10^{-2}$) & Logarithmic &  &  \\
& Gas-phase metallicity & $Z_g\ /\ \mathrm{Z_\odot}$ & (0.00355, 3.55) & Logarithmic & & \\

\hline
Calibration & Zero order & $P_0$ & (0.75, 1.25) & Gaussian & $\mu = 1$ & $\sigma$ = 0.1 \\
& First order & $P_1$ & ($-0.25$, 0.25) & Gaussian & $\mu = 0$ & $\sigma$ = 0.1\\
& Second order & $P_2$ & ($-0.25$, 0.25) & Gaussian & $\mu = 0$ & $\sigma$ = 0.1\\
\hline
Noise & White noise scaling & $a$ & (0.1, 10)  & logarithmic & & \\
\hline
\end{tabular}
\endgroup
\label{table:params}
\end{sidewaystable*}
\clearpage}

\afterpage{
\begin{table*}
\caption{Posterior percentiles for the 22 free parameters of the Bagpipes model we fit to our spectroscopic and photometric data. Full definitions of these parameters, along with their associated prior distributions, are given in Extended Data Table \ref{table:params}.}
\begingroup
\setlength{\tabcolsep}{5pt}
\renewcommand{\arraystretch}{1.1}
\begin{tabular}{llll}
\hline
Parameter & 16\textsuperscript{th} Percentile & 50\textsuperscript{th} percentile & 84\textsuperscript{th} Percentile \\
\hline
 $z$ & 4.6580 & 4.6582 & 4.6584 \\
 $\sigma$ / km s$^{-1}$ & 256 & 269  & 283 \\
\hline
 log$_{10}(M_*\ /\ \mathrm{M_\odot})$ & 10.81 & 10.83 & 10.85 \\
  $Z_*\ /\ \mathrm{Z_\odot}$ & 0.09 & 0.11 & 0.12 \\
  $\alpha$  & 42 & 155 & 511 \\
  $\beta$  & 10 & 22 & 113 \\
  $\tau$ / Gyr  & 0.71 & 0.75 & 0.79 \\
\hline
  $A_V$ / mag & 0.00 & 0.02 & 0.04 \\
  $\delta$ & $-0.05$ & 0.00 & 0.05 \\
  $B$   & 0.81 & 2.47 & 4.15 \\
  $\epsilon$ & 1.78 & 2.98 & 4.22 \\
\hline

  $\alpha_{\lambda < 5000\mathrm{\angstrom}}$   & $-0.79$ & $-0.52$  & $-0.27$ \\
 $\alpha_{\lambda > 5000\mathrm{\angstrom}}$  & 0.14 & 0.40 & 0.72 \\
 $f_{\mathrm{H}\alpha\mathrm{,\,broad}}$ / erg s$^{-1}$ cm$^{-2}$ & $2.8\times10^{-20}$& $3.4\times10^{-20}$ & $3.9\times10^{-20}$ \\
 $\sigma_{\mathrm{H}\alpha\mathrm{,\,broad}}$ / km s$^{-1}$  & 4100 & 4400 & 4700 \\
 $f_{5100}$ / erg s$^{-1}$ cm$^{-2}$ \AA$^{-1}$  & $1.18\times10^{-17}$ & $1.26\times10^{-17}$ & $1.35\times10^{-17}$ \\

\hline

log$_{10}(U)$  & -3.65 & -3.00 & -2.37 \\
 $Z_g\ /\ \mathrm{Z_\odot}$ & 0.02 & 0.17 & 1.54  \\

\hline
 $P_0$  & 1.13  & 1.15 & 1.18 \\
 $P_1$  & 0.09 & 0.1 & 0.12 \\
 $P_2$  & $-0.08$ & $-0.07$ & $-0.06$ \\
\hline
$a$  & 1.49 & 1.51 & 1.53 \\
\hline
\end{tabular}
\endgroup
\label{table:params_values}
\end{table*}
\clearpage}

\setcounter{figure}{0}

\afterpage{
\begin{figure}
	\includegraphics[width=\columnwidth]{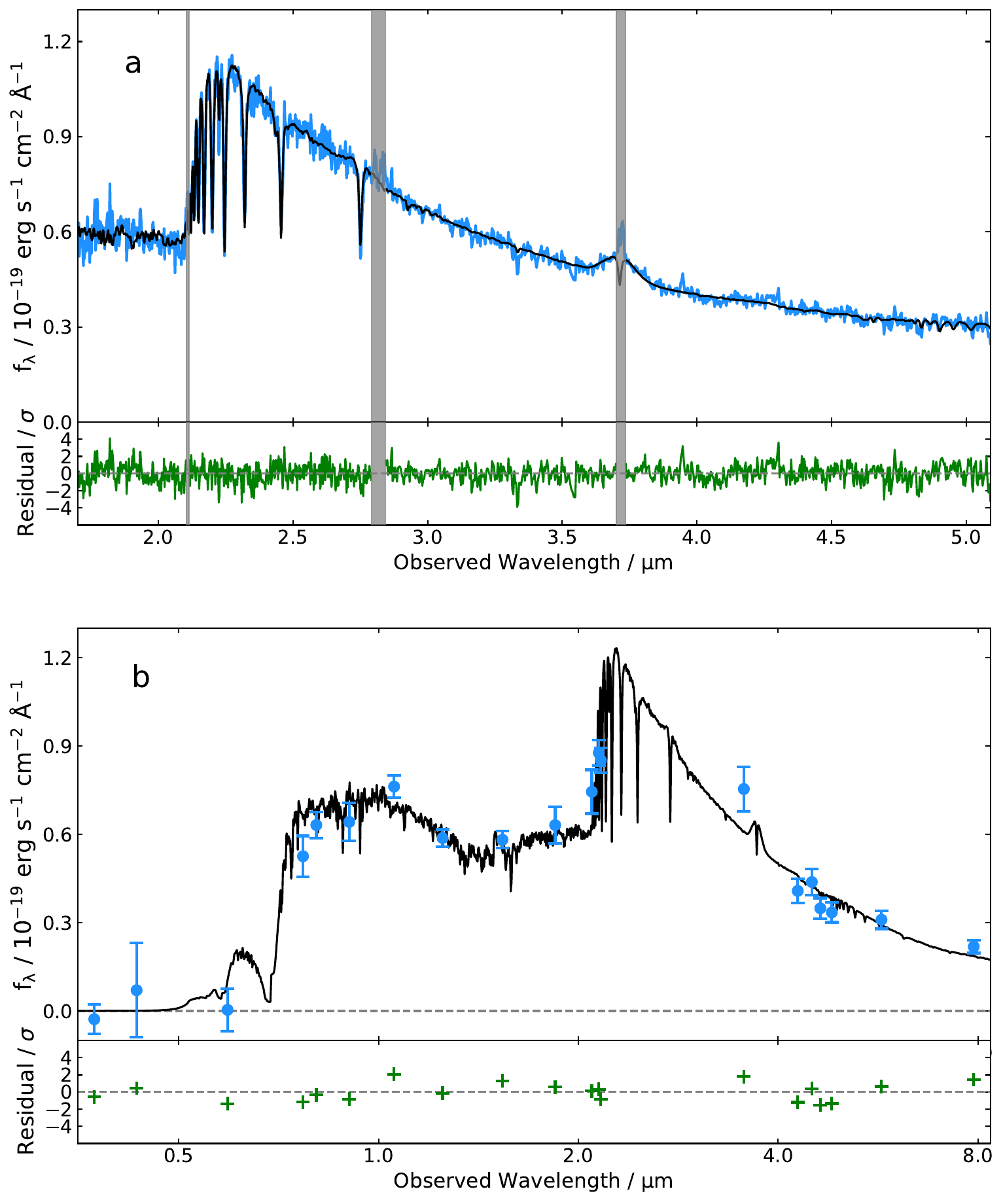}
    \caption{Model fit to our full combined dataset. Panel a shows our spectroscopic dataset in blue, with the full Bagpipes fitted model shown in black, and residuals between model and data shown below in green. Regions of our spectroscopic dataset masked during our Bagpipes fitting are shaded gray. Panel b shows our photometric data, again in blue, with the full model again shown in black, and residuals below marked with green crosses.}
    \label{fig:spectrum3}
\end{figure}
\clearpage}

\afterpage{
\begin{figure}
	\includegraphics[width=\columnwidth]{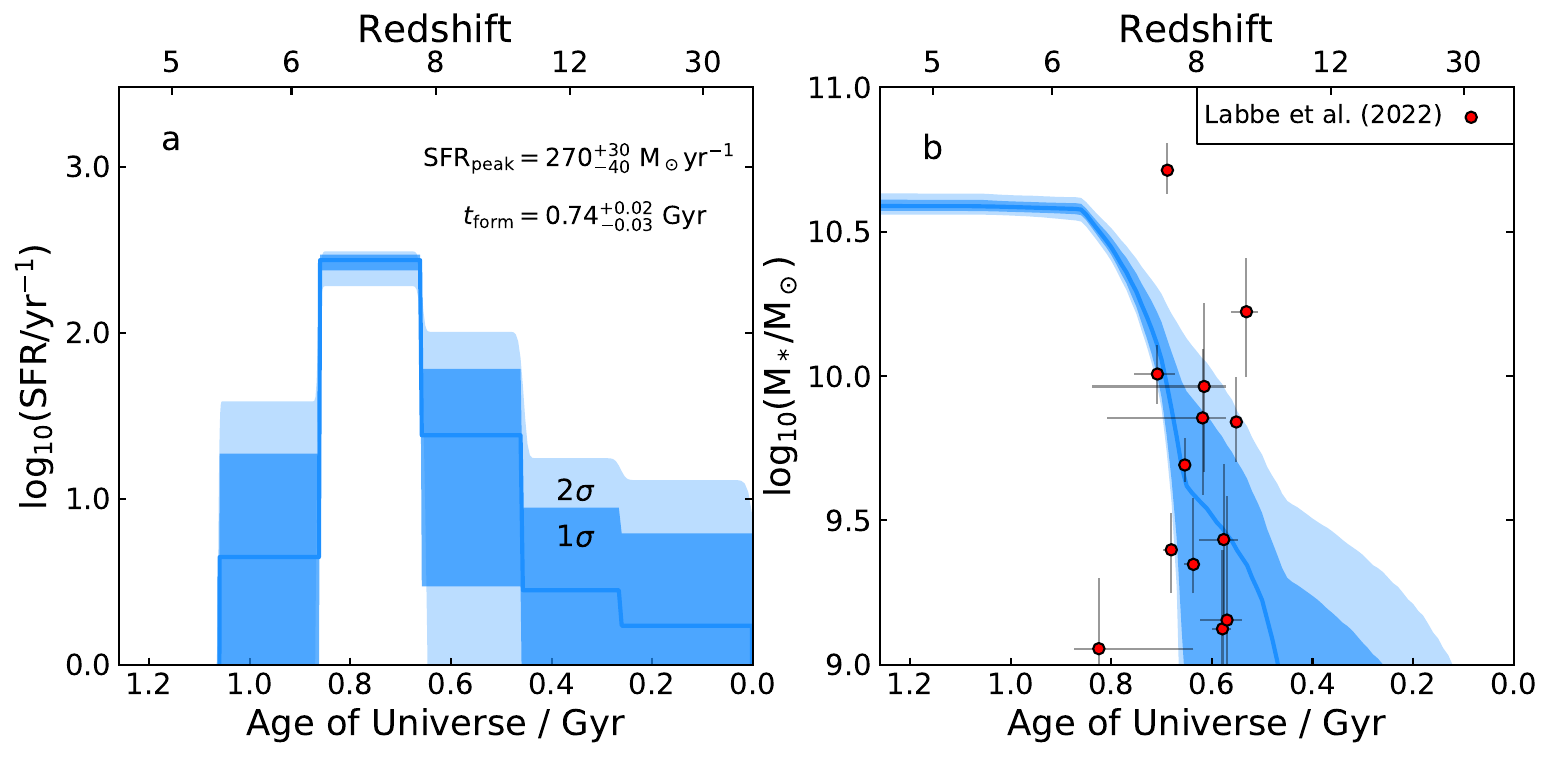}
 	\includegraphics[width=\columnwidth]{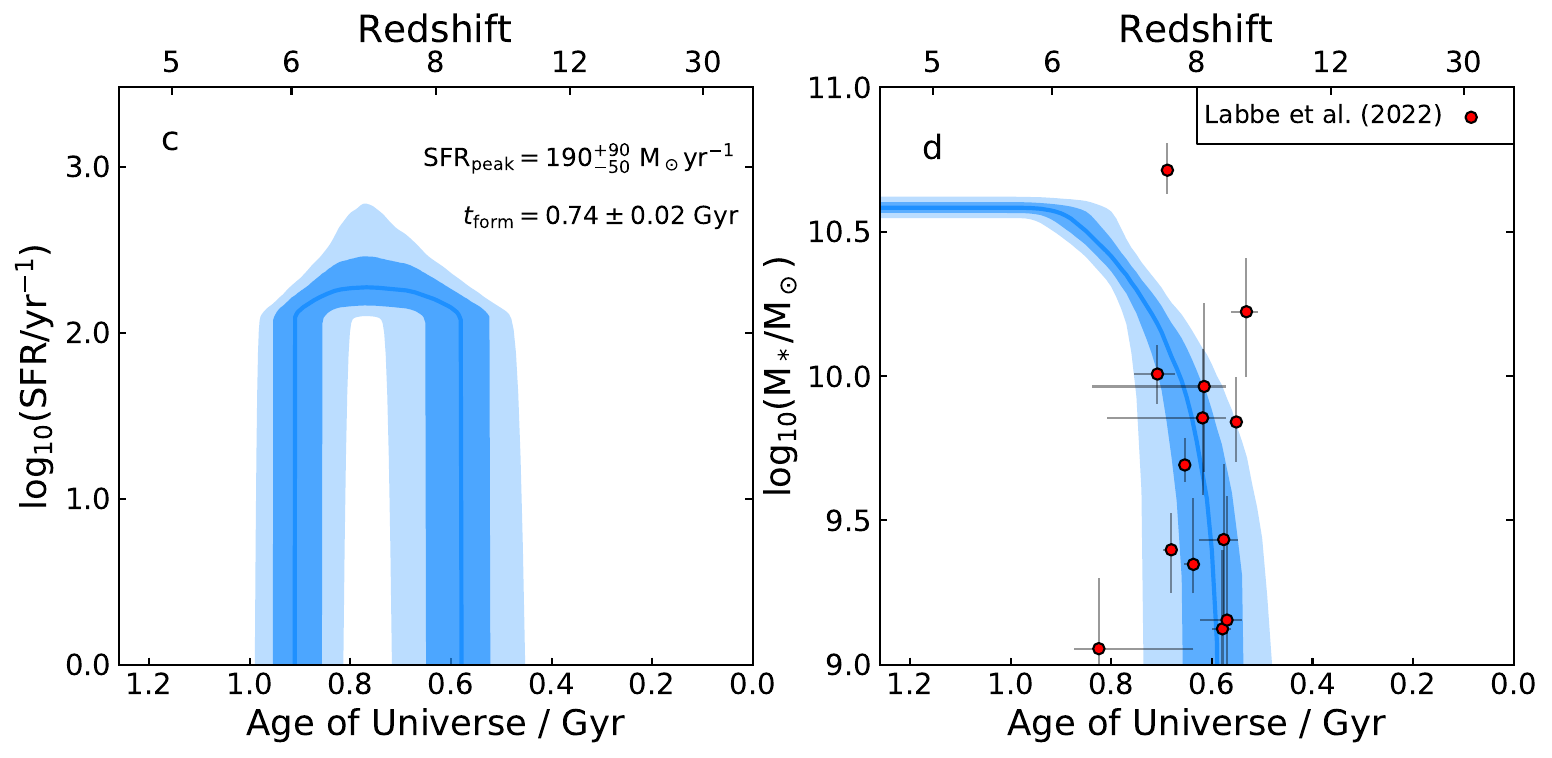}
    \caption{The star-formation rate and stellar mass of GS-9209 as a function of time under the assumption of different models. Panels a and b show the star-formation rate and stellar mass respectively as a function of time under the assumption of the continuity non-parametric SFH model \cite{Leja2019a} (as an alternative to the double-power-law model used in the main analysis and shown in Figure \ref{fig:sfh}). Panels c and d again show the star-formation rate and stellar mass respectively as a function of time, this time assuming a top-hat SFH model. Consistent formation times are recovered using all three models.}
    \label{fig:alt_sfhs}
\end{figure}
\clearpage}

\end{document}